\documentclass[pra,preprint,showpacs]{revtex4-2}
\usepackage[centertags]{amsmath}
\usepackage[utf8]{inputenc}
\usepackage{amsfonts}
\usepackage{rotating}
\usepackage{amssymb}
\usepackage{amsthm}
\usepackage{amsmath}
\usepackage{newlfont}
\usepackage{epsfig}
\usepackage{amscd}
\usepackage{graphicx}
\usepackage{epsfig}
\usepackage{footnote}
\usepackage{lipsum}
\usepackage{color}
\usepackage{xcolor}
\usepackage{graphicx}
\usepackage{multirow}
\usepackage{enumerate}
\usepackage{caption}
\usepackage{subcaption}
\usepackage{multirow}
\usepackage{amscd}

\newcommand{\beq}{\begin{equation}}
\newcommand{\eeq}{\end{equation}}
\newcommand{\ba}{\begin{array}}
\newcommand{\ea}{\end{array}}
\newcommand{\bea}{\begin{eqnarray}}
\newcommand{\eea}{\end{eqnarray}}
\begin{document}

\begin{center}
{\large \sc \bf  { Teleportation via spin-1/2 chain in solid-state quantum architecture}
}

\vskip 15pt

{\large
 E.B.~Fel'dman,  S.I. Doronin,  E.I.~Kuznetsova, A.I.~Zenchuk$^{*}$
}

\vskip 8pt

{\it 
Federal Research Center of Problems of Chemical Physics and Medicinal Chemistry RAS,
Chernogolovka, Moscow reg., 142432, Russia}.

{\it $^*$Corresponding author. E-mail:  zenchuk@itp.ac.ru}

\vskip 8pt

\end{center}


\begin{abstract}
We propose the protocol for preparing the maximally entangled  Bell state between remote qubits at the ends of the spin-1/2 chain governed by the specially engineered  nearest-neighbor XX-Hamiltonian with excited central spin as the initial state. This method does not require including optical constituent in the teleportation protocol and  can be implemented in the quantum devices with solid-state architecture for teleporting unknown states or  organizing quantum gates between remote qubits. {A superconducting flux-qubit chain is  an example of such devises.}
 \end{abstract}


\maketitle

\section{Introduction}

Quantum teleportation allows to transfer the unknown one-qubit pure quantum  state between remote qubits  traditionally called Alice (A) and Bob (B), provided that those qubits  share  an entangled state \cite{Werner_1989,BBCJPW,Popescu,HHH}.  This is  a phenomenon whose realizability is completely based on  the quantum entanglement. 
First realization of   teleportation in laboratory over meter-scale distance is based on optical photons  \cite{BPMEWZ,FSBFKP}.
Then, quantum teleportation over long distances of over 143 kilometers and distant teleportation using ground-to-satellite communication was 
realized in experiments with photons in Refs. \cite{MHSWK} and \cite{Retal} respectively. 
The short-distance teleportation between ions in the same trap was performed in \cite{RHRHB,BCSBIJLLOW}.
In  \cite{OMMHDM}, teleportation  between atoms in different traps separated by a meter distance is demonstrated  where entanglement is established  by a beam of entangled photons.
Teleportation over 3-meter distance between  diamond spin qubits  is implemented  in 
\cite{PHBDBT} via installing entanglement between electron spins.
Teleportation in networks is implemented in \cite{HPBBBH}.
Teleportation between qubits of distant superconducting chips is considered in \cite{QLNHWZHCLLZDY}.  Significant attention attracts the quantum gate teleportation \cite{MDNANSAL,JTSL,MRRBMDK}, which allows realization of quantum gates involving  remote qubits.
In all proposed methods  of distant  teleportation the entanglement between  remote  ions or atoms is established via intermediate photons that are emitted by those ions (atoms), interfere and are detected. 

In our  teleportation algorithm we avoid implementation of the optical technique {for entangling two distant qubits}, which  simplifies the technical aspect of  teleportation over relatively short communication distance. The   entanglement between the remote  qubits is established   via the spin chain connecting them, which is applicable in solid-state quantum devises to organize teleportation between qubits of the same quantum platform.   

\section{General algorithm}

First of all, we show that the Hamiltonian dynamics with certain initial state results in the maximal entangled Bell state between two end-qubits.  
To this end, we  consider the inhomogeneous odd-node symmetrical spin-1/2 chain where the first and last spins are  marked as $A$ and $B$ and other qubits form the transmission line $TL$.
We assume that the dynamics is governed by the Hamiltonian preserving the excitation number, which allows us to work in the one-excitation subspace of the state space of a quantum system. 

Let  $|n\rangle$ denote the quantum state  where the $n$th  qubit is excited and others are in the ground states.
To create the desired entanglement between the qubits $A$  and $B$ we consider the centrally-symmetric spin-1/2 chain and  the initial state with the middle excited spin:
\begin{eqnarray}\label{Psi000}
|\Psi(0)\rangle = |(N+1)/2\rangle.
\end{eqnarray}
Then, due to the chain symmetry,  we can select the maximally entangled state of the qubits $A$ and $B$ from the superposition evolution state  as follows:
\begin{eqnarray}\label{Psi00}
|\Psi(t)\rangle &=& e^{-i Ht } |{(N+1)/2}\rangle = \alpha(t) (|10\rangle_{AB} + |01\rangle_{AB})\otimes |0\rangle_{TL} + \beta(t) |g(t)\rangle,\\\nonumber
&&2 |\alpha|^2 + |\beta|^2 =1,  \;\; \langle g|\Big((|10\rangle_{AB} + |01\rangle_{AB})\otimes |0\rangle_{TL}\Big) =0, 
\end{eqnarray}
where $(|10\rangle_{AB} + |01\rangle_{AB})\otimes |0\rangle_{TL}\equiv
|1\rangle_A |0\rangle_{TL} |0\rangle_B + |0\rangle_A |0\rangle_{TL} |1\rangle_B$.
We are interested  in such Hamiltonian, that can provide the condition  $|\alpha(t_0)|^2 =\frac{1}{2}$ at some time instant $t_0$.  In this case $\beta(t_0)=0$ and the state (\ref{Psi00}) reads
\begin{eqnarray}\label{Psi0}
|\Psi(t_0)\rangle = \frac{e^{i \varphi}}{\sqrt{2}} (|10\rangle_{AB} + |01\rangle_{AB})\otimes |0\rangle_{TL} ,
\end{eqnarray}
where $\varphi$ is the  phase of $\alpha(t_0)$. Thus, the qubits $A$ and $B$ are maximally entangled.

At the time instant $t_0$, we  start the teleportation assuming that all succeeding operations are  fast in comparison with $\|H\|^{-1}$, where $\|\cdot \|$ is some norm. 
Thus, teleporting the arbitrary state of  the additional qubit $\tilde A$ (which was not involved into the evolution)   
\begin{eqnarray}\label{varphi}
|\varphi\rangle_{\tilde A} = a|0\rangle_{\tilde A} +b|1\rangle_{\tilde A}, \;\;|a|^2+|b|^2=1,
\end{eqnarray}
we  start with the following state of the whole system  at $t_0$:
\begin{eqnarray}\label{st_tel}
|\Phi_1(t_0)\rangle =  \frac{e^{i \varphi}}{\sqrt{2}}  |\varphi\rangle_{\tilde A}  (|10\rangle_{AB} + |01\rangle_{AB})\otimes |0\rangle_{TL} 
\end{eqnarray}
{and evaluate the standard teleportation algorithm.}
To teleport the two-qubit unitary transformation $U$ \cite{MDNANSAL}, i.e., to create the operator  $U_{\tilde A \tilde B}$ starting with $U_{\tilde A A}$ and $U_{\tilde B B}$, where  $\tilde B$ is an additional qubit  (which is not included into the above Hamiltonian dynamics, similar to  the qubit $\tilde A$) in the state  $|\chi\rangle_{\tilde B}$, we  consider the state 
\begin{eqnarray}\label{U_tel}
|\Phi_1(t_0)\rangle =  \frac{e^{i \varphi}}{\sqrt{2}} U_{\tilde A A} U_{\tilde B B}|\varphi\rangle_{\tilde A}  (|10\rangle_{AB} + |01\rangle_{AB})\otimes |0\rangle_{TL}  |\chi\rangle_{\tilde B}.
\end{eqnarray}
After that,  the usual  gate-teleportation algorithms can be implemented  which is described in the above quoted references and is  not reproduced here.

According to the above discussion, the key step of teleportation is creating the maximally  entangled state  (\ref{Psi0}) of remote qubits $A$ and $B$  via the protocol 
proposed below.

\section{Maximal entanglement via engineered spin-1/2 chain}

  The qubits $A$ and $B$ are connected by the transmission line $TL$ of  $N-2$ qubits, so that the total number of qubits in the chain is $N$.
 In order to the spin evolution lead to the Bell state of  qubits $A$ and $B$ we use the odd-node spin chain governed by  the nearest-neighbor $XX$-Hamiltonian 
 \begin{eqnarray}\label{Ham}
 H=\sum_{j=1}^{N-1} D_j (I_{j,x}  I_{j+1,x} + I_{j,y}  I_{j+1,y}), \;\; [H,I_z]=0, 
 \end{eqnarray}
 where $D_{j}$ are  the dipole-dipole coupling constants to be determined, $I_{j,\alpha}$, $\alpha=\{x,y,z\}$, are the $\alpha$-projections of the $j$th spin, $I_z=\sum_{j=1}^N I_{j,z}$. 

 We impose  the  two following symmetries on the coupling constants:
 \begin{eqnarray}\label{sym1}
 &&D_{N-i} = D_i, \;\; i=1,\dots,\frac{N-1}{2} , \\\label{sym2} 
 &&D_{(N-1)/2 -k+1} = D_{k},\;\;k=2,\dots, \lfloor(N-1)/4\rfloor.
 \end{eqnarray}
 Thus $D_1$ and $D_{(N-1)/2}$ of the first half and $D_{N-1}$, $D_{(N+1)/2}$ of the second half of the chain are not subjected to the symmetry (\ref{sym2}). Relation between nodes in each of these pairs will be obtained below in (\ref{Dk3}).
 
Let us write the system of equations for the eigenvectors of the Hamiltonian $H$:
\begin{eqnarray}\label{ev}
H u_k = \lambda_k u_k,\;\;k=1,\dots,N.
\end{eqnarray}
In components, this system reads
\begin{eqnarray}\label{u1}
&&D_1 u_{2,k}  = \lambda_k u_{1,k},\\\label{uj}
&&D_{j-1} u_{j-1,k} + D_j u_{j+1,k} = \lambda_k u_{j,k},\;\;j=2,\dots,N-1,\\\label{uN}
&&D_{1} u_{N-1,k} = \lambda_N u_{N,k}.
\end{eqnarray}
For creating the Bell state at the time instant $t_0$ with the initial state $|\Psi(0)\rangle$ given in Eq. (\ref{Psi000}), we need to satisfy the condition 
\begin{eqnarray}\label{1UN}
\left|\langle 1| e^{-i Ht} |{(N+1)/2}\rangle\right|^2= \left|\langle N| e^{-i Ht} |{(N+1)/2}\rangle\right|^2=    \frac{1}{2}.
\end{eqnarray}
Let $U=( u_1,\dots, u_N)$. Since, according to (\ref{ev}),  $H=U\Lambda U^\dagger$ with $\Lambda = {\mbox{diag}}(\lambda_1,\dots,\lambda_N)$,  Eq.(\ref{1UN}) yields
\begin{eqnarray}\label{1Uksum}
\left|  \sum_{k=1}^N u_{1,k} e^{- i\lambda_k t} u_{(N+1)/2,k}\right|^2=\frac{1}{2}.
\end{eqnarray}  
It seamed out that not all eigenvectors $u_k$ are included into the sum in Eq.(\ref{1Uksum}).
To prove this statement, first of all we note that symmetry (\ref{sym1}) prompts us the  two following reductions for $u_{j,k}$:
\begin{eqnarray}
\label{sym11}
&&u_{(N+1)/2 +j ,k} = u_{(N+1)/2 -j ,k},\\\label{sym12}
&&u_{(N+1)/2 +j ,k} =- u_{(N+1)/2 -j ,k},\;\; j=1,\dots,(N+1)/2-1 .
\end{eqnarray}
These reductions separate the set of the eigenvectors $u_k$, $k=1,\dots N$, in two families
and reduce the number of  independent equations in system (\ref{u1})-(\ref{uN}) 
to the system of 
$(N+1)/2$  equations for each $k$:
\begin{eqnarray}\label{u12}
&&D_1 u_{2,k}  = \lambda_k u_{1,k},\\\label{uj2}
&&D_{j-1} u_{j-1,k} + D_j u_{j+1,k} = \lambda_k u_{j,k},\;\;j=2,\dots,(N-1)/2,\\\label{uN2}
&&D_{(N-1)/2} u_{(N-1)/2,k} \pm D_{(N-1)/2} u_{(N-1)/2,k} = \lambda_k u_{(N+1)/2,k},\\\nonumber
&&k=1,\dots,N,
\end{eqnarray}
the signs ``$+$'' and  ``$-$''  in Eq.(\ref{uN2}) correspond to, respectively, reductions (\ref{sym11}) and (\ref{sym12}).
Obviously, the left hand side of Eq. (\ref{uN2}) under  reduction (\ref{sym12}) (sign ``$-$'') equals zero. Then the right hand side of this equation yields
\begin{eqnarray}
 u_{(N+1)/2,k} =0. 
\end{eqnarray}
This means, that the sum in (\ref{1Uksum}) includes only terms with eigenvectors satisfying reduction (\ref{sym11}). Therefore, we concentrate on this reduction hereafter. In this case, Eq. (\ref{uN2}) gets the form
\begin{eqnarray}\label{uN22}
2D_{(N-1)/2} u_{(N-1)/2,k}  = \lambda_k u_{(N+1)/2,k},\;\;
k=1,\dots,N.
\end{eqnarray}

 As the next step, 
we show that system (\ref{u12}), (\ref{uj2}) and (\ref{uN22}) can be considered as an eigensystem problem for some  $(N+1)/2 \times  (N+1)/2 $ Hamiltonian. 
 For this purpose, we first note that the
 eigenvectors $u_k$ are normalized:
 \begin{eqnarray}\label{norm1}
 2\sum_{j=1}^{(N-1)/2} u_{j,k}^2 + u_{(N+1)/2,k}^2=1, \;\; \forall k.
 \end{eqnarray}
 Let us introduce the following change of eigenvectors:
 \begin{eqnarray}\label{utu}
 u_{j,k} =\frac{ \hat u_{j,k}}{\sqrt{2}},\;\; j =1,\dots, (N-1)/2,\;\; u_{(N+1)/2,k}= \hat u_{(N+1)/2,k}
 \end{eqnarray}
 with normalization 
 \begin{eqnarray}
 \sum_{j=1}^{(N-1)/2} \hat u_{j,k}^2 + \hat u_{(N+1)/2,k}^2=1,
 \end{eqnarray}
 which follows from the normalization (\ref{norm1}). 
 If now we  relate  the coupling constants $D_{(N-1)/2}$ and $D_1$ by
  \begin{eqnarray}\label{Dk}
\label{Dk3}
D_{(N-1)/2}=D_1/\sqrt{2},
 \end{eqnarray}
then system (\ref{u12}), (\ref{uj2},  (\ref{uN22}) gets the form
 \begin{eqnarray}\label{2u12}
&&D_1 \hat u_{2,k}  = \lambda_k \hat u_{1,k},\\\label{2uj2}
&&D_{j-1} \hat u_{j-1,k} + D_j \hat u_{j+1,k} = \lambda_k \hat u_{j,k},\;\;j=2,\dots,(N-1)/2,\\\label{2uN2}
&&D_{1} \hat u_{(N-1)/2,k}  = \lambda_k  \hat  u_{(N+1)/2,k},\\\nonumber
&&k=1,\dots,(N+1)/2.
\end{eqnarray}
 The system (\ref{2u12}) -  (\ref{2uN2})  in view of symmetry (\ref{sym2}) is the eigensystem  problem for the  nearest-neighbor  XX-Hamiltonian $\tilde H$ governing the dynamics of the symmetric  $(N+1)/2$-node  spin chain:
 \begin{eqnarray}\label{H2}
 \tilde H = \sum_{j=1}^{(N+1)/2-1} D_j  (I_{j,x}  I_{j+1,x} + I_{j,y}  I_{j+1,y}).
 \end{eqnarray}
 The problem of perfect state transfer (PST) between the end nodes of this chain has been considered in set of papers, for instance,  in Refs.\cite{CDEL,KS}. We use the choose of $D_i$ given in 
 \cite{CDEL}:
\begin{eqnarray} \label{Dk}
&&D_k=\frac{\mu}{2}\sqrt{k\Big(\frac{N+1}{2}-k\Big)}
,\;\; k=1,\dots, (N-3)/2\\\label{Dk2}
&& D_{(N-1)/2}=\frac{\mu}{2\sqrt{2}}\sqrt{\frac{N-1}{2}},
 \end{eqnarray}
 where $\mu$ is some positive parameter.
 Then, the probability amplitude of the end-to-end  excited state transfer along the  chain of $(N+1)/2$ spins governed by the Hamiltonian (\ref{H2})  is
 \begin{eqnarray}\label{prob}
 \langle 1|\sum_{k=1}^{(N+1)/2}  \hat u_{k} e^{-i \lambda_k t} \hat u^T_k |(N+1)/2\rangle =
 \sum_{k=1}^{(N+1)/2}  \hat u_{1,k} e^{-i \lambda_k t }\hat u_{(N+1)/2,k} = \left(-i \sin \frac{\mu}{2} t\right)^{(N-1)/2}.
 \end{eqnarray}
 At $\mu t_0=\pi$, the probability of the end-to-end  state transfer is one. 
 If we go back to the functions $u_{j,k}$ via Eq.(\ref{utu}), then Eq.(\ref{prob}) gets the form:
  \begin{eqnarray}\label{prob2}
  \sum_{k=1}^{(N+1)/2}  u_{1,k} e^{-i \lambda_k t } u_{(N+1)/2,k} =\frac{1}{\sqrt{2}}  \left(-i \sin \frac{\mu}{2} t\right)^{(N-1)/2}.
 \end{eqnarray}
 Thus, at $\mu t_0=\pi$, the probability of the excited state transfer from the center spin of the original $N$-node chain  to the  first one is $\frac{1}{2}$.
 Due to the symmetry, the probability of the excited state transfer to the $N$th spin is also $\frac{1}{2}$. This means that the state of the whole system of $N$ spins  is Eq.(\ref{Psi0}) with $e^{i \varphi} = (-i)^{(N-1)/2}$, which includes the Bell-state of the end nodes $A$ and $B$ and the ground states of the all other nodes.
 In this way we create the maximally entangled state of the two-end-qubits $A$ and $B$ at the time instant $t_0=\pi/\mu$, which does not depend on the chain length $N$. 
 
 {
 \section{Teleportation via superconducting qubit chain}
 Teleportation  over 6 mm distance between two superconducting qubits coupled by the waveguide resonator was considered in \cite{SSOKBLEPFW}. In this case, the entangled Bell state can be created by the local operations of controlled-phase gate and $y$-rotations. To create the Bell state between remote qubits, flying micro-wave photons can be used \cite{QLHWNZH} which allows to teleport an arbitrary state over 64 meters.

 The proposed algorithm of entangled state creation based on qubit-dynamics is an alternative to the above ones. It can be implemented in superconducting quantum platforms where the approximate  Hamiltonian governing the evolution of flux-qubits (neglecting interaction with environment)  reads  \cite{MREKTA} $H=\frac{1}{2}\sum_{j=1}^N \Delta_j \sigma^{(z)}_j + \frac{1}{2} \sum_{i<j, i,j=1}^N g_{ij} (\sigma^{(x)}_i \sigma^{(x)}_j + \sigma^{(y)}_i \sigma^{(y)}_j)$, where $\sigma^{(\alpha)}_j$, $\alpha=x,y,z$, are the Pauli matrices, $\Delta_j$ are the energy gaps between states $|0\rangle$ and $|1\rangle$ of the $j$th qubit, $g_{ij}$ are the coupling constants, which  are  tunable via a special microwave-driving tool \cite{NHYNLT}. 
 The nearest-neighbor approximation of this  Hamiltonian with equal  $\Delta_j=\Delta$  reduces (in the rotation frame of references) to Hamiltonian  (\ref{Ham})  
 with the coupling constants $g_{j,j+1}\equiv D_j$ tunable to  form (\ref{Dk}), (\ref{Dk2}).
 The excitation of the middle qubit, required for creating the desired entangled state (\ref{Psi0}), 
can be achieved via relaxing all the qubits to the ground state by cooling  and then applying a resonant microwave $\pi$-pulse
 carried by a microwave line connected to the middle qubit \cite{CW,CNHM}. Formally, this $\pi$-pulse corresponds to applying the $\sigma^{(x)}$-operator to the middle qubit. Note that the flux-qubit  dynamics with different initial conditions along the superconducting chain was considered in \cite{RZPL} for  Duke model.

{
Realization of considered Hamiltonian evolution  requires that the maximal  creatable value of  the coupling constant $g^{max}$ is bigger than the maximal coupling constant  in Eqs.(\ref{Dk}), (\ref{Dk2}): $g^{max} \geq D^{max} \equiv D_{\lfloor (N-1)/4\rfloor}\stackrel{N\gg 1}{\approx} \frac{\mu N}{8}$.  For the fixed $\mu$, this relation yields the upper  boundary for $N$: $N^{max} = 8 g^{max}/\mu = 8g^{max} t_0/\pi$, i.e., $N^{max}$  linearly grows with $t_0$ which is the same for all   $N$ if $N\le N^{max}$ and coupling constants are constructed according to Eqs.(\ref{Dk}), (\ref{Dk2}). For instance, if we set $\mu =10^{4} Hz$, then $t_0 =\pi/\mu \approx  3\times  10^{-4}$ and, using  $g^{max}= 730 MHz$ given in  \cite{MREKTA}  (Table A.1.), we obtain  $N^{max}=584000$. 

We also have to mention the direct method for creating the  entanglement between remote spins via SWAP operators, when the first qubit ($A$) is initially entangled with the nearest neighbor and then SWAP operators  step-by-step  move away the second qubit of the entangled state till the required last qubit $B$:
\begin{eqnarray}
 \frac{1}{\sqrt{2}} \prod_{i=2}^{N-1} SWAP_{i,i+1}(|01\rangle + |10\rangle)|\underbrace{0\dots 0}_{N-2}\rangle =
\frac{1}{\sqrt{2}} (|01\rangle_{AB}  + |10\rangle_{AB}) |0\rangle_{TL},
\end{eqnarray}
where $TL$ includes all $N-2$  qubits between $A$ and $B$, $SWAP_{i,j}$ swaps the states of the $i$th and $j$th qubits. 
  In this case, the time needed to establish the entanglement between the qubits $A$ and $B$ linearly increases with $N$. In the best case, $t_0 \sim \frac{N}{g^{max}}$, which is similar to the estimation of $t_0$ in the evolution case.  {In addition,   entangling  remote qubits via evolution method  does not require  involving the manual manipulations,} while  the $SWAP$-method requires special ordering of SWAP-operators which is an additional technical complication.

Nevertheless, the   method based on the micro-wave photons   remains the only one  to establish  entanglement between the qubits of different platforms. }
 
 {We  emphasize   that the proposed algorithm represents a mathematical model that includes the basic idea of teleportation completely based on the solid-state architecture. Of course, perturbations of  coupling constants effect on PST reducing the state-transfer fidelity \cite{ZASO,ZASO2} and therefore reducing the fidelity of teleportation.
The noise in quantum platform as well as the approximate nature of the driving  Hamiltonian  create obstacles for perfect deterministic teleportation.
 In this respect, we shall note that the  probabilistic  method of PST considered in Ref.\cite{FWZ_arxive2025}  is less sensitive to the Hamiltonian perturbations and  is applicable to any Hamiltonian preserving the excitation number in the system.  That  method is based on the measurement of some additional qubit (ancilla) at the time instant $t_0$.
 }}

 {\subsection{Effects of Hamiltonian perturbation}
  In this section we present  the robustness   analysis of entangled-state creation and  teleportation algorithm with respect to  perturbations of coupling constants in the Hamiltonian.
  
First, we briefly recall the basic steps of the  teleportation algorithm using  the  state $|\Phi_1(t_0)\rangle $ in Eq.(\ref{st_tel}) setting  zero the unimportant common phase $\varphi$ \cite{Werner_1989,BBCJPW,Popescu,HHH}. 
\begin{enumerate}
\item
Apply the CNOT operator $C_{\tilde AA} = |1\rangle_{\tilde A} \, _{\tilde A}\langle 1| \otimes \sigma^{(x)}_A +  |0\rangle_{\tilde A} \, _{\tilde A}\langle 0| \otimes I_A $ to the qubits $\tilde A$ and $A$  and 
  the Hadamard operator $H_{\tilde A}$ to $\tilde A$, 
\begin{eqnarray}\label{Phi1}
&&|\Phi_2\rangle =H_{\tilde A}C_{\tilde AA} |\Phi_1(t_0)\rangle   = 
\frac{1}{2} \Big(|00\rangle_{\tilde A A} ( a|1\rangle_{B} + b |0\rangle_{B}) +
|01\rangle_{\tilde A A} ( a|0\rangle_{B} + b |1\rangle_{B})  +\\\nonumber
&&
|10\rangle_{\tilde A A} ( a|1\rangle_{B} - b |0\rangle_{B}) +
|11\rangle_{\tilde A A} ( a|0\rangle_{B} - b |1\rangle_{B})\Big) |0\rangle_{TL}.
\end{eqnarray}
\item
Measure the states of the qubits $\tilde A$ and $A$ obtaining one of four possible states    $|\psi\rangle_{\tilde AA}$, the result of measurement is classically sent to  Bob. 
\item
After getting this result, Bob  knows the   state $|\phi\rangle_{B}$ of $B$, see Table \ref{Table:1}. Then, applying the appropriate unitary transformation from the last column of Table \ref{Table:1} to $B$, Bob finishes  the state teleportation.
\begin{table}[h!]
\begin{tabular}{|ccc|}
\hline
measured state,               & state of $ B$,  & operators applied   \cr
$|\psi\rangle_{\tilde AA}$& $|\phi\rangle_{ B}$ &to $B$, $U_B$\cr
\hline
$|00\rangle_{\tilde AA} $& $ \frac{1}{\sqrt{2}} ( a|1\rangle_{B} + b |0\rangle_{B})$&$\sigma^{(x)}_B$\cr
$|01\rangle_{\tilde AA} $ & $\frac{1}{\sqrt{2}} ( a|0\rangle_{B} + b |1\rangle_{B})$&$I_B$ \cr
$|10\rangle_{\tilde AA} $& $ \frac{1}{\sqrt{2}} ( a|1\rangle_{B} - b |0\rangle_{B})$&$\sigma^{(z)}_B \sigma^{(x)}_B$\cr
$|11\rangle_{\tilde AA} $ & $\frac{1}{\sqrt{2}} ( a|0\rangle_{B} - b |1\rangle_{B})$&$\sigma^{(z)}_B$ \cr
\hline
\end{tabular}
\caption{ {The results of measurement of qubits $\tilde A$ and $A$, the appropriate state of $B$ and the operators to be applied to $B$ }}
\label{Table:1}
\end{table}
\end{enumerate}

If the coupling constants deviate from the prescribed values (\ref{Dk}), (\ref{Dk2}), 
\begin{eqnarray}\label{Ddk}
D_k \to D_k +\mu \varepsilon d_k, \;\;\varepsilon\ll 1,\;\;-1/2\le d_k \le 1/2,
\end{eqnarray}
then the state of the whole quantum system at the time instant of state registration ($\tau_0 =\mu t_0 =\pi$ in the perfect case) is not the product state of the form (\ref{Psi0}). 
This state can be written as
 \begin{eqnarray}\label{Psid}
 &&
 |\Psi(t_0)\rangle  = \Big(c_1 |10\rangle_{AB} +c_2  |01\rangle_{AB}\Big) |0\rangle_{TL} + \delta(\varepsilon) |00\rangle_{AB}|\chi\rangle_{TL} ,\\\label{norm}
 &&
  |c_1|^2 + |c_2|^2 + |\delta|^2=1,
  \;\; \delta\ll 1,
 \end{eqnarray}
 where $|\chi\rangle_{TL}$ is a superposition of 1-excitation states of the subsystem $TL$. 
The deformation 
 $|\chi\rangle_{TL}$ is normalized and orthogonal to  $|0\rangle_{TL}$:  $_{TL}\langle \chi|\chi\rangle_{TL} =1$, $_{TL}\langle 0|\chi\rangle_{TL} =0$.
 Then, attaching the additional qubit  $\tilde A$ in the arbitrary state  $|\varphi\rangle_{\tilde A}$ to be teleported (see Eq. (\ref{varphi})) and performing step 1 of the  teleportation algorithm given in the beginning of this subsection we obtain, instead of (\ref{Phi1}),
 \begin{eqnarray}\label{Phi12}
&&|\Phi_2\rangle =H_{\tilde A}C_{\tilde AA} |\varphi\rangle_{\tilde A} |\Psi(t_0)\rangle   = \\\nonumber
&&
\frac{1}{\sqrt{2}} \Big(|00\rangle_{\tilde A A} ( ac_2|1\rangle_{B} + b c_1 |0\rangle_{B}) +
|01\rangle_{\tilde A A} ( ac_1|0\rangle_{B} + bc_2 |1\rangle_{B})  +\\\nonumber
&&
|10\rangle_{\tilde A A} ( ac_2|1\rangle_{B} - bc_1 |0\rangle_{B}) +
|11\rangle_{\tilde A A} ( ac_1|0\rangle_{B} - bc_2 |1\rangle_{B})\Big) |0\rangle_{TL} +
\\\nonumber
&&\frac{\delta}{\sqrt{2}}\Big( a |00\rangle_{\tilde AA}+ b |01\rangle_{\tilde AA} + a |10\rangle_{\tilde AA}- b |11\rangle_{\tilde AA}     \Big)|0\rangle_{B} |\chi\rangle_{TL} .
\end{eqnarray}
 After steps 2 and 3 of the teleportation algorithm, the state of  the system reads
 \begin{eqnarray}
|\Phi_3\rangle_{TL,B}&=&\left\{
\begin{array}{ll}
  G_1^{-1}\Big(|0\rangle_{TL}  (a c_2 |0\rangle_B + b c_1 |1\rangle_B)+\delta a |\chi\rangle_{TL} |1\rangle_B \Big) &  {\mbox{measured st.}}\;\; |00\rangle_{\tilde AA} \cr
  G_2^{-1}\Big(|0\rangle_{TL}  (a c_1 |0\rangle_B + b c_2 |1\rangle_B)+\delta b|\chi\rangle_{TL} |0\rangle_B  \Big) & {\mbox{measured st.}}\;\; |01\rangle_{\tilde AA} \cr
  G_1^{-1}\Big(|0\rangle_{TL}  (a c_2 |0\rangle_B + b c_1 |1\rangle_B) -\delta a |\chi\rangle_{TL} |1\rangle_B \Big) &  {\mbox{measured st.}}\;\; |10\rangle_{\tilde AA} \cr
  G_2^{-1}\Big(|0\rangle_{TL}  (a c_1 |0\rangle_B + b c_2 |1\rangle_B)-\delta b |\chi\rangle_{TL} |0\rangle_B  \Big) & {\mbox{measured st.}}\;\; |11\rangle_{\tilde AA}
\end{array}\right.\\
\nonumber
&&G_1=\sqrt{|c_1|^2(|b|^2-|a|^2) + |a|^2} , \;\; G_1=\sqrt{|c_1|^2(|a|^2-|b|^2) + |b|^2}.
 \end{eqnarray}
 We see that the state   $|\varphi\rangle_{\tilde A}$ is not perfectly teleported to the qubit $B$ and the 
teleportation  fidelity is given by 
 \begin{eqnarray}\label{f}
f&=&\Big|\big(_{TL}\langle 0|\, _B\langle \varphi|\big)   | \Phi_3\rangle_{TL,B}  \Big|^2 \\\nonumber
&&=\left\{ 
\begin{array}{ll}
G_1^{-2}\Big| |a|^2c_2 + |b|^2c_1\Big|^2  &  {\mbox{if measured state is}}\;\; |00\rangle_{\tilde AA} \;\;{\mbox{or}}\;\;  |10\rangle_{\tilde AA}  \cr
G_2^{-2}\Big | |a|^2c_1 + |b|^2c_2\Big|^2 &  {\mbox{if measured state is}}\;\; |01\rangle_{\tilde AA} \;\;{\mbox{or}}\;\; |11\rangle_{\tilde AA}
  \end{array}
    \right. .
 \end{eqnarray}
In addition, we can refer to the method of increasing the fidelity  of state transfer based on special manipulating with end-nodes coupling constants  \cite{GKMT,ABCVV}, which  is  useful  for  optimizing  the  particularly realized spin chain. 

\subsubsection{Numerical simulations}
Now we present results on numerical simulation of the algorithm entangling remote qubits. 
We  consider  perturbation of the coupling constants (\ref{Dk}), (\ref{Dk2}) in  form (\ref{Ddk})
where $d_k$ are random real numbers inside of  the interval  $-1/2 \le d_k \le 1/2$. 
As characteristics, we consider the probability $|\delta|^2$, describing   the impact of perturbation on the maximally entangled state of $A$ and $B$,  and the fidelity $F$ averaged over the Bloch sphere (setting $a=\cos \theta$, $b= e^{i \phi} \sin \theta$, $0\le \phi < 2\pi$, $0\le \theta \le \pi$):
\begin{eqnarray}\label{Favr}
F=\frac{1}{2} \int_{0}^\pi f(\theta) \sin \theta d \theta.
\end{eqnarray}
Operating with $f$, we consider only  the upper line in  the right hand side of (\ref{f}), because using the lower line yields much the same result.

In numerical experiment, we perform 500 simulations with different sets of randomly fixed $d_k$ in (\ref{Ddk}). The perturbation amplitude  $\varepsilon$ and the number of qubits in the chain $N$ take the values 
\begin{eqnarray}\label{pertampl}
&&\varepsilon = 10^{-j}, \;\;j= 0, 1, \dots, 6,\\\label{chainN}
&&N=11, 101, 1001, 10001.
\end{eqnarray}
On the graphs in Fig.\ref{Fig:e}a and Fig.\ref{Fig:e}b, we show, respectively, the quantities $\log_{10}  |\overline{\delta}|^2$ and  $\log_{10}(1-\overline{F})$ as  functions of $\log_{10} \varepsilon$, 
where $|\overline{\delta}|^2$ and $\overline{F}$ are, respectively,   $|\delta|^2$  and $F$ averaged over 500 random set of perturbations $d_k$ of the coupling constants  in  (\ref{Ddk}) for a fixed  value of the  chain length $N$ from set (\ref{chainN}), $N$ increases downwards.  The obtained results (red circles)  are   approximated by the lines 
\begin{eqnarray}\label{line1}
\log_{10} |\overline{\delta}|^2 = \varkappa_\delta \log_{10} \varepsilon +C_\delta,   \;\; 
\log_{10} (1-\overline{F} )= \varkappa_F \log_{10} \varepsilon +C_F
\end{eqnarray}
 with some parameters $\varkappa_\delta$, $C_{\delta}$, $\varkappa_F$, $C_F$ collected in  Table \ref{Table:2}. 
On the graphs in Fig.\ref{Fig:ee}a and Fig.\ref{Fig:ee}b, we show the same quantities $\log_{10} |\overline{\delta}|^2$ and  $\log_{10}(1-\overline{F})$ as  functions of $\log_{10} N$ for a fixed value of the perturbation amplitude $\varepsilon$ from set (\ref{pertampl}), $\varepsilon$ decreases downwards. 
 The obtained results (red circles)  are   approximated by the lines
 \begin{eqnarray}\label{line2}
 \log_{10} |\overline{\delta}|^2 = \tilde \varkappa_\delta \log_{10} N +\tilde C_\delta,\;\;   \log_{10}(1-\overline{F}) =\tilde  \varkappa_F \log_{10} N+\tilde C_F.
 \end{eqnarray} 
Parameters $\tilde \varkappa_\delta$, $\tilde C_{\delta}$, $\tilde\varkappa_F$, $\tilde C_F$ of the approximating lines  are collected in  Table \ref{Table:2}. 
\begin{figure}[ht]
\begin{subfigure}{0.45\textwidth}
    \includegraphics[width=0.7\textwidth]{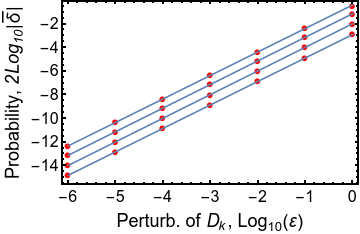}
    \caption{}
    \end{subfigure}
    \begin{subfigure}{0.45\textwidth}
      \includegraphics[width=0.7\textwidth]{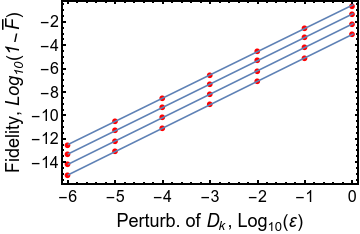}
      \caption{}
    \end{subfigure} 
    \caption{(a)  $\log_{10} |\delta|^2$   and (b)  $\log_{10}(1-\overline{F})$  as functions of  $\log_{10} \varepsilon$ for different  numbers of qubits in the chain $N=11, 101, 1001, 10001$, increasing downwards.  The probability $|\overline{\delta}|^2$ decreases while  the fidelity $\overline{F}$ increases with a decrease in $\varepsilon$ at fixed $N$.  }
\label{Fig:e}
\end{figure}
\begin{figure}[ht]
\begin{subfigure}{0.45\textwidth}
    \includegraphics[width=0.7\textwidth]{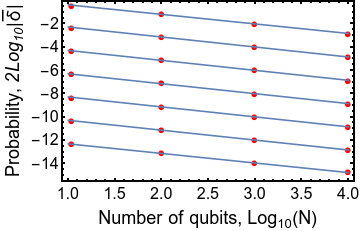}
    \caption{}
    \end{subfigure}
    \begin{subfigure}{0.45\textwidth}
      \includegraphics[width=0.7\textwidth]{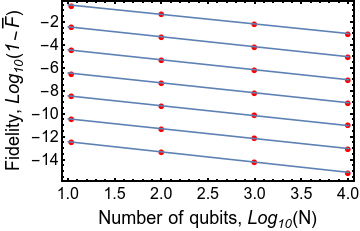}
      \caption{}
    \end{subfigure} 
    \caption{(a)  $\log_{10} |\overline{\delta}|^2$  and (b)  $\log_{10}(1-\overline{F})$  as functions of  $\log_{10} N$ for different  perturbation amplitudes $\varepsilon = 10^{-j}$, $j=0,\dots,6$, decreasing  downwards. The probability   $|\overline{\delta}|^2$  decreases,  while the fidelity  $\overline{F}$ increases with an  increase  in  $N$ at fixed $\varepsilon$.
     }
\label{Fig:ee}
\end{figure}
  \begin{table}
 \begin{tabular}{|c|cc|cc|}
 \hline
 &\multicolumn{2}{|c|}{Fig.\ref{Fig:e}a} &  \multicolumn{2}{|c|}{Fig.\ref{Fig:e}b} \cr
 \hline
$N$ & $\varkappa$ & $C$ & $\varkappa$ & $C$ \cr
\hline
11 &1.988&-0.381&1.989& -0.503\cr
101&1.998&-1.116&  1.998&  -1.262\cr
1001&2.000&-1.964& 2.003& -2.115\cr
10001&1.986&-2.881&2.005& -3.007\cr
\hline
 \end{tabular}
 \begin{tabular}{|c|cc|cc|}
 \hline
 &\multicolumn{2}{|c|}{Fig.\ref{Fig:ee}a} &  \multicolumn{2}{|c|}{Fig.\ref{Fig:ee}b} \cr
 \hline
$\varepsilon$ & $\varkappa$ & $C$ & $\varkappa$ & $C$ \cr
\hline
1 & -0.820& 0.460& -0.833&0.349\cr
$10^{-2}$&-0.855& -3.418& -0.868& -3.533\cr
$10^{-4}$& -0.856&-7.411& -0.864& -7.537\cr
$10^{-6}$&-0.821&-11.473& -0.889&-11.502\cr
\hline
 \end{tabular}
 \caption{The parameters of approximating lines  in Fig.\ref{Fig:e} and Fig.\ref{Fig:ee}. The $\varkappa$-parameters confirm that the lines on each plate are essentially parallel. 
 }
 \label{Table:2}
 \end{table}
 
 According to (\ref{line1}), (\ref{line2}) and  Table \ref{Table:2}, 
 $|\overline{\delta}|^2\approx 10^{C_\delta} \varepsilon^{1.99} \approx 10^{\tilde  C_\delta} N^{-0.84}$,
 $1-\overline{F}\approx  10^{C_F} \varepsilon^{2.00}\approx 10^{\tilde C_F} N^{-0.86}$, here we average the  slopes $\varkappa$ over the data in Table \ref{Table:2}. 
 We can underline several properties of relations among $\delta$, $\varepsilon$ and $N$.
 \begin{enumerate}
 \item
 The probability $|\overline{\delta}|^2$ is quadratic function of the perturbation amplitude $\varepsilon$  because  
 $\delta \sim \varepsilon$ in  the evolution governed by the nearest-neighbor XX-Hamiltonian.  
 \item
 As a consequence of n.1,  Eq.(\ref{f}) and Eq.(\ref{norm}), $1-\overline{F}$ is a linear  function of $|\overline{\delta}|^2$ and therefore a quadratic function of $\varepsilon$.
 \item
The probability  $|\overline{\delta}|^2$ is almost linear function of $N^{-1}$. This is caused by the  fact that  the minimal coupling constant   in (\ref{Dk2}) $D_{min}\sim \sqrt{N}$, therefore  the relative perturbation amplitude $\varepsilon/D_{min}$ decreases with $N$ as $\sim1/ \sqrt{N}$ at
  fixed $\varepsilon$.  The same behavior exhibits $|\delta|$: $|\delta|\sim1/\sqrt{N}$,  i.e.,  $|\delta|^2\sim N^{-1}$.
 \item
As a consequence of n.3, Eq.(\ref{f}) and Eq.(\ref{norm}),   $1-\overline{F}$ is almost linear function of $N^{-1}$ at fixed $\varepsilon$.
 In other words, the effect of any fixed $\varepsilon$ vanishes as $N\to\infty$. 
 \end{enumerate}
 The performed numerical analysis demonstrates that our algorithm is rather   robust  with respect to the  perturbations of coupling constants and therefore might be useful in applications.
 }
 
 \section{Conclusions}
We present the algorithm allowing to establish the maximal entanglement between remote qubits connected by a chain of spin-1/2 particles governed by the nearest-neighbor XX-
Hamiltonian  with specially engineered  coupling constants. This method  allows to avoid the including optical technique to organize the remote entanglement over short distance and  can be implemented in the solid-state quantum architecture. Such model can be  useful, for instance,  for teleporting a quantum state between remote qubits of the same quantum platform which can be applied to perform the data-exchange between quantum circuits.
Another implementation of remote entangled qubits is teleportation of quantum gates to perform the unitary operation between remote qubits on a quantum platform. We propose only one realization of the symmetric chain establishing the entanglement between remote qubits using coupling constants (\ref{Dk}). However, other realizations of maximal end-to-end  entanglement are also possible and have been obtained in numerical simulations.  We also note that maximal entanglement between the qubits $A$  and $B$ allows the deterministic perfect teleportation. On the contrary, the non-maximal entanglement between the qubits $A$  and $B$, caused by, for instance, small  perturbations of the Hamiltonian decreases fidelity of the teleported state. 

{We discuss the possible application of the proposed teleportation algorithm  to the flux-qubit chain on the superconducting platform. {The long-distance entanglement in such  platform can be established via 
micro-wave photons. Our algorithm is  an alternative to the existing one and can be more effective for entangling qubits separated by intermediate-scale distance.  Using    solely solid basis for teleportation allows to simplify the technical part of teleportation between remote qubits.    
We present the  analysis of the effect of the  Hamiltonian perturbations on the entangling algorithm and demonstrate that perturbations reduce the fidelity of teleported state. However, our algorithm is rather tolerant  to such perturbations which is confirmed by numerical simulations showing that deviation of the fidelity from 1, i.e.,  the quantity $1-F$, is proportional to square of the perturbation amplitude $\varepsilon$. 
We consider that the teleportation (more precisely, the teleportation of  unitary transformations) based on our algorithm  can be effectively used for realization of quantum operations between remote qubits  inside of single quantum platform. }

{\bf  Acknowledgments.}  The work is funded as a state task of Russian Fundamental Investigations (State Registration No. 124013000760-0). 

{\bf CONFLICT OF INTEREST} Authors have no conflict of interest.

\end{document}